# Collective Rabi-driven vibrational activation in molecular polaritons

Carlos M. Bustamante*[1], Franco P. Bonafé[1], Richard Richardson[2], Michael Ruggenthaler[1], Wenxiang Ying[3], Abraham Nitzan[3], Maxim Sukharev[2,4], and Angel Rubio[1,5]

[1]Max Planck Institute for the Structure and Dynamics of Matter and Center for Free-Electron Laser Science, Luruper Chaussee 149, Hamburg 22761, Germany
[2]Department of Physics, Arizona State University, Tempe, Arizona 85287, United States
[3]Department of Chemistry, University of Pennsylvania, Philadelphia, Pennsylvania 19104, United States
[4]College of Integrative Sciences and Arts, Arizona State University, Mesa, Arizona 85212, United States
[5]Initiative for Computational Catalysis (ICC), Flatiron Institute, Simons Foundation, 162 5th Avenue, New York, NY 10010 USA

*carlos.bustamante@mpsd.mpg.de

**Abstract**

Molecular polaritons arise from electronic or vibrational strong coupling (ESC and VSC) with confined electromagnetic fields. While these have been widely studied, the influence of electron-nuclear dynamics in driven cavities remains largely unknown. Here, we report a previously unrecognized mechanism of vibrational activation that emerges under collective ESC in driven optical cavities. Using simulations that self-consistently combine Maxwell's equations with quantum molecular dynamics, we show that collective electronic Rabi oscillations coherently drive nuclear motion. This effect is captured using both vibrational wave-packet dynamics in a minimal two-level model and atomistic simulations based on time-dependent density-functional tight-binding theory. Vibrational activation depends non-monotonically on the Rabi frequency and is maximized when the collective polaritonic splitting resonates with a molecular vibrational mode. The mechanism exhibits features consistent with a stimulated Raman-like relaxation mechanism. Our predictions are robust under realistic cavity conditions and provide the conditions in which they could be verified experimentally.

Inside optical cavities, molecules can interact strongly with confined electromagnetic modes, forming hybrid light–matter states that can modify molecular properties, a phenomenon underlying polaritonic chemistry [1–12]. However, not all cavity-induced effects involve real photons or excited-state hybridization, which we refer to as *polaritons*. We therefore adopt the term *endyons* for polaron-like quasiparticles arising from vacuum electromagnetic fluctuations that induce static, zero-point renormalizations [13], distinguishing them from polaritons. This distinction is essential, as the present work focuses on the nonequilibrium dynamics of *molecular polaritons*, rather than on ground-state reactivity in "dark" cavities.

Strong light–matter coupling can involve either electronic or vibrational molecular transitions. Electronic strong coupling (ESC) enables control of excited-state dynamics and photochemical pathways [14–23]. Vibrational strong coupling (VSC), by contrast, involves infrared-active modes and has been widely explored for its potential to modify ground-state chemistry without external driving [24–29]. Most theoretical studies of vibro-polaritons rely on *ab initio* ground-state calculations to predict cavity-modified vibrational spectra [30–32], limiting their ability to describe driven vibronic dynamics and leaving the interplay between ESC, cavity fields, and nuclear motion largely unexplored.

In this work, we demonstrate that individual molecular vibrational modes can be coherently activated inside driven optical cavities through ESC, and that this activation is governed by the Rabi oscillations of the coupled light–matter system. By studying realistic cavity geometries, we establish a direct connection between electronic polaritonic dynamics and nuclear motion, revealing a mechanism for selectively driving vibrations under nonequilibrium conditions. To uncover this effect, we employ semiclassical simulations that self-consistently combine Maxwell's equations for the cavity electromagnetic fields with quantum descriptions of molecular electron dynamics [33, 34]. As a minimal model, we first adopt the Born–Oppenheimer (BO) approximation and represent the molecules using two potential energy surfaces, on which we propagate coupled vibrational wavepackets. This approach, referred to throughout the text as the two-level model, provides a description of the vibrational response in the presence of electronic Rabi oscillations.

To treat realistic polyatomic molecules, we further use a multiscale framework combining finite-difference time-domain (FDTD) solutions of Maxwell's equations with atomistic electronic dynamics at the density functional tight-binding level [33]. In this implementation, the electronic density matrix of each molecule is propagated self-consistently using the DFTB+ package [35], while nuclear motion is treated classically within the Ehrenfest approximation [36]. Maxwell's equations are solved in one and two spatial dimensions (see Methods section), whereas the molecular systems are described in full three dimensions, allowing a realistic representation of all cavity modes, spatial field profiles, and metal mirrors with frequency-dependent dielectric response [37].

Despite the known limitations of Ehrenfest dynamics for nuclear motion [38], we show that this level of theory captures the essential features of the driven vibrational activation induced by ESC. The results point to a stimulated Raman-like process driven by the interplay of upper and lower polaritonic fields as the underlying mechanism. From the perspective of the electromagnetic response, this behavior is consistent with polariton relaxation mediated by phonon emission. The vibrational driving frequency is determined by the collective Rabi splitting of the molecular ensemble rather than single-molecule coupling, and the efficiency and selectivity of the process depend on cavity mode structure and quality factor. This collective character places the effect squarely within the realm of molecular polariton physics and highlights the role of cavity mediated intermolecular interactions, and cavity modified electronic interaction, in shaping nuclear dynamics[39–41].

## Rabi-driven vibrational activation via polaritonic resonance

We describe coupled electronic–nuclear dynamics within the Born–Oppenheimer (BO) approximation using two potential energy surfaces along a nuclear coordinate (Fig. 1A), taken as the bond length of a diatomic molecule.

To simulate the dynamics of this system under electronic strong coupling, we propagate Schrödinger's equation using the Hamiltonian defined in Eq. 6 of the Methods section, self-consistently coupled to one-dimensional Maxwell's equations. The system consists of a realistic Fabry–Perót (FP) cavity formed by two 40-nm-thick gold mirrors with a Drude–Lorentz dielectric response, separated by 305.5 nm. An ensemble of molecules is placed at the center of the cavity on individual grid points, spanning a region 20 nm thick. The molecular electronic transition is tuned to be in resonance with the first cavity mode. The strength of the light–matter coupling is controlled by varying the electronic transition dipole moment $\mu_{eg}$. Additional details of the simulation setup are provided in the Methods section (*Two level model with vibrations*).

Following excitation by a 5-fs pump pulse, ESC induces repeated absorption and emission, producing Rabi oscillations until cavity losses dissipate the field. The characteristic frequency of this process, referred to as the Rabi frequency ($\Omega$), can be extracted from the transmission spectra by measuring the separation between the upper polariton (UP) and lower polariton (LP) peaks. When vibrational degrees of freedom are included, the UP peak splits into multiple vibro-polaritonic features following Franck–Condon selection rules (Fig. 1B) [42]. This prevents the assignment of a single well-defined Rabi frequency. Nevertheless, the overall increase of the effective Rabi splitting with increasing $\mu_{eg}$ remains evident from the growing LP-UP

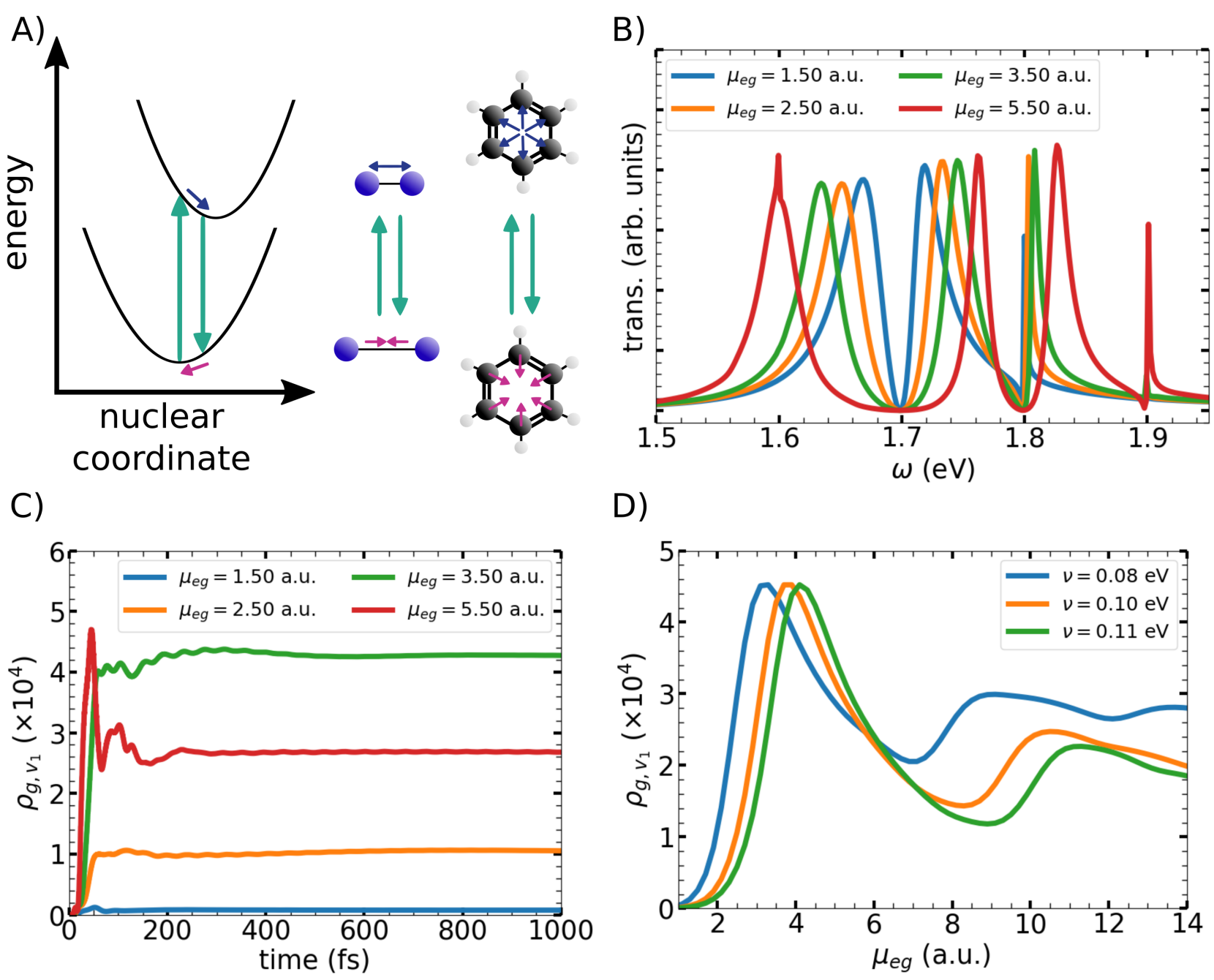

Figure 1: A) Schematic illustration of vibrational excitation induced by electronic strong coupling (ESC). The two parabolic curves represent the ground- and excited-state potential energy surfaces of a generic molecule. Cyan arrows indicate electronic absorption and emission processes, while blue and pink arrows denote the associated nuclear displacements during electronic excitation and relaxation. A diatomic molecule and a benzene molecule are shown as representative examples. B) Calculated transmission spectra for the two-level model as the effective electronic transition dipole moment $\mu_{eg}$ is varied, illustrating the evolution of the polaritonic response under increasing light–matter coupling. The spectra show the LP and UP peaks below and above 1.7 eV, respectively. Due to the quantum vibrational degrees of freedom, the UP peak is split. [42]. C) Time evolution of the occupation of the first vibrational state $v_1$ of the electronic ground state for the two-level model with vibrational frequency $\nu = 0.1, \mathrm{eV}$. D) Steady-state occupation of the first vibrational state $v_1$ of the electronic ground state as a function of $\mu_{eg}$ for three different vibrational frequencies. For clarity, the blue and green curves are normalized to the maximum of the orange curve.

separation.

Electronic Rabi oscillations induce vibrational activation, which we track by the time evolution of the occupation of the first vibrational state ($v_1$) of the electronic ground state, shown in Fig. 1C. Following an initial transient, the occupation of this vibrational state reaches a steady value once optical energy is dissipated by cavity losses. This final occupation depends on $\mu_{eg}$ and thus on the effective Rabi frequency. Because the vibrational mode is optically inactive, further relaxation is inefficient.

Plotting the steady-state vibrational population versus $\mu_{eg}$ reveals a pronounced maximum (Fig. 1D). We attribute this feature to a resonance between the vibrational frequency of the ground state and the Rabi frequency associated with ESC. Consistent with this interpretation, increasing the vibrational frequency $\nu$, shifts the resonance to larger $\mu_{eg}$.

From a semiclassical perspective, this behavior can be understood as a driven damped harmonic oscillator, where the driving force arises from the time-dependent excited-state electronic density, analogous to displacive excitation of coherent phonons [43]. The periodicity of this force is set by the electronic Rabi oscillations, whose frequency depends on the light–matter coupling strength. Energy transfer to the vibrational degree of freedom is maximized when the driving frequency resonates with the molecular vibrational frequency, resulting in a maximum occupation of the first vibrational state.

In addition to the dominant resonance, Fig. 1D reveals a secondary maximum at larger values of the electronic transition dipole moment. Analysis of the underlying vibrational populations indicates that this feature originates from a Raman-type transition between the $v = 0$ and $v = 2$ vibrational states of the electronic ground state. A detailed investigation of this higher-order process and its dependence on system parameters lies beyond the scope of the present work.

For the semiclassical approximation the damping leads to a broadening of the resonant response. In the present simulations, losses at the cavity mirrors effectively introduce damping into the system. Other electron–phonon dephasing mechanisms are not considered here, but they would not alter the conclusions presented below. Within this framework, shorter cavity lifetimes, corresponding to stronger damping, are expected to produce vibrational activation over a broader range of values of $\mu_{eg}$ or, equivalently, Rabi frequencies. The two-level model captures this behavior, which shows broader, less defined resonances with increasing mirror losses (Supplementary Fig. 1).

Increasing the size of the molecular slab shifts the resonance to lower $\mu_{eg}$ values (Supplementary Fig. 2) and reduces the population at resonance. While this may suggest a dilution of vibrational activation in large ensembles, such a conclusion is premature and requires further investigation.

A final feature of the Rabi-driven vibrational activation is that, under resonant conditions, the vibrational population scales with the fourth power of the driving field amplitude (Supplementary Fig. 3), consistent with a Raman-like excitation mechanism (see Discussion).

## Rabi-driven vibrational activation in benzene within a spatially structured cavity

We extend the analysis to polyatomic molecules in a FP cavity using Ehrenfest dynamics to describe coupled electronic and nuclear motion. We begin the study with benzene, whose first optically allowed $\pi \rightarrow \pi^*$ excitation weakens the C–C bonds and leads to an expanded excited state ring geometry[44].

Under resonant cavity conditions and following excitation, the light-matter energy exchange induces a periodic contraction and expansion of the benzene ring, analogous to the diatomic molecule case (Fig. 1A). Our simulations employ a one-dimensional Maxwell solver with two 50-nm aluminium mirrors (Drude–Lorentz dielectric response [37]) separated by 298 nm. The third cavity mode is tuned to the first electronic transition of benzene at 6.79 eV, predicted by DFTB and consistent with TDDFT [45].

An ensemble of 201 benzene molecules (each lying in the $xy$ plane) is placed over a 200 nm region at the cavity center to probe spatial effects. Further details of the simulation setup are provided in the Methods section (*DFTB systems*). The strength of the light–matter coupling is varied by changing the effective molecular concentration parameter $N_M$ in Eq. 5, from $1.69 \times 10^{-3}$ nm$^{-3}$ to $3.74 \times 10^{-2}$ nm$^{-3}$ (molar concentrations of 0.0028 M and 0.0621 M).

Vibrational activation is quantified by the time-averaged vibrational potential energy (VPE) of each normal mode, obtained by projecting nuclear displacements onto normal modes and averaging after cavity

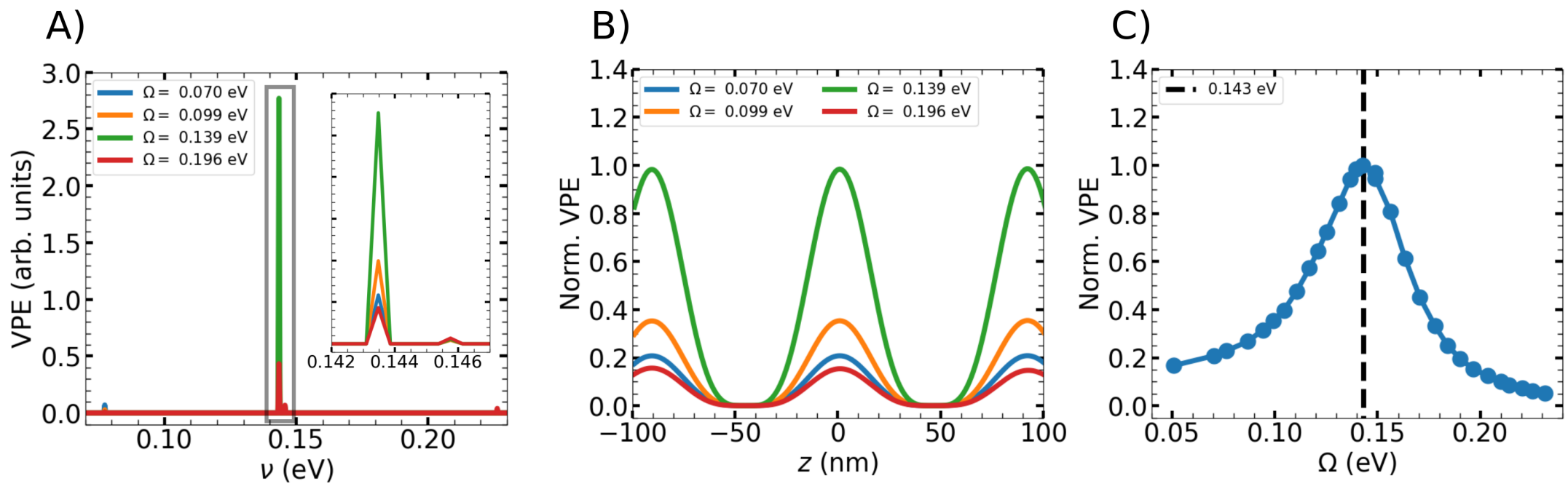

Figure 2: A) Vibrational potential energy (VPE) per normal mode of the central benzene molecule ($z = 0.0, \mathrm{nm}$), plotted as a function of vibrational frequency and averaged over the last 100 fs of the simulation for different values of the Rabi splitting. The inset provides a magnified view of the highlighted frequency range. B) VPE of the benzene breathing mode ($\nu = 0.143, \mathrm{eV}$) for each molecule inside the cavity, shown as a function of molecular position for different values of the Rabi splitting. The VPE maps out the spatial profile of the third cavity mode, i.e., with two zeros at about ± 49 nm. C) Normalized VPE of the benzene breathing mode of the central molecule as a function of the Rabi splitting extracted from the transmission spectra, obtained by varying the molecular density parameter $N_M$.

energy dissipation, for more details see Methods section (*Average vibrational potential energy*).

The VPE is plotted as a function of the vibrational frequencies of the benzene molecule in Fig. 2A, yielding a representation analogous to a vibrational spectrum, for different values of the Rabi frequency. Owing to the classical treatment of the nuclear motion, the Rabi frequency can be determined unambiguously from the LP-UP splitting in transmission spectra (Supplementary Fig. 4).

Only the breathing mode (0.143 eV) shows strong activation; while others show weak or no response. Its nuclear dynamics closely follows the behavior described previously and illustrated schematically in Fig. 1A. This selectivity reflects both the frequency mismatch between the other modes and the applied Rabi splittings, as well as the small projection of Rabi-induced nuclear motion onto them. Symmetry arguments may further rationalize this behavior [46], and will be explored in future work.

As demonstrated in our previous work [33], the interaction between cavity modes and molecules depends strongly on spatial position of the molecule due to the structure of the electromagnetic field. This spatial dependence is directly reflected in the Rabi-driven vibrational activation. As shown in Fig. 2B, the breathing-mode VPE is maximized near field antinode regions and suppressed near nodes.

The resonant character of the Rabi-driven vibrational activation is further illustrated in Fig. 2C, where the VPE of the breathing mode shows a maximum when the Rabi frequency matches the vibrational frequency, corresponding to the condition $\Omega = \nu$. The Supplementary Movie visualizes the corresponding field, nuclear and electronic dynamics of the central molecule under resonant conditions. Tuning away from resonance progressively suppresses the vibrational activation, and no higher-order processes predicted by the two-level model are observed within the parameter range explored here.

Extending the analysis to two-dimensional cavities (Supplementary Fig. 5A) increases losses, reducing mode selectivity. In benzene, this leads to simultaneous activation of multiple modes for a given Rabi splitting (Supplementary Fig. 5B), broader resonances, and smaller VPE variations (Supplementary Fig. 5C). Thus, it is shown that Rabi-driven vibrational activation is robust to cavity dimensionality, but vibrational mode selectivity requires high-quality cavities, potentially achievable by tailoring the cavity geometry using inverse design strategies [34].

# Rabi-driven multimode vibrational activation

In the previous examples, Rabi-driven nuclear motion projects mainly onto a single vibrational mode. This behavior, however, should not be regarded as universal. In more complex molecular systems, the electronic excitation can couple to multiple vibrational degrees of freedom. To illustrate such situations, we consider the pentacene molecule, which provides a suitable balance between structural complexity and analytical tractability due to its symmetry.

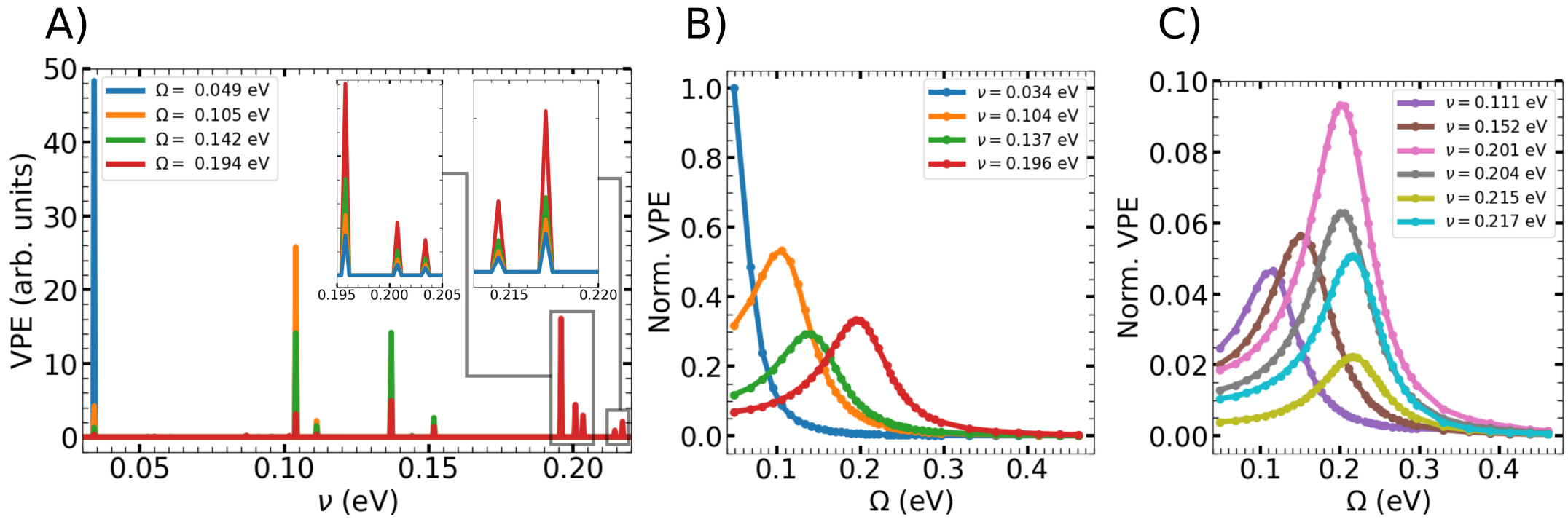

Figure 3: A) VPE of the central pentacene molecule plotted as a function of vibrational frequency, averaged over the final 100 fs of the simulation for different values of the Rabi frequency. The insets provide magnified views of the highlighted frequency ranges. B) and C) show the normalized VPE of the most strongly affected vibrational modes of the central pentacene molecule plotted as a function of the Rabi frequency. Both panels share the same axis scale.

For these simulations, we employ a one-dimensional cavity geometry consisting of two 50-nm aluminium mirrors with a Drude–Lorentz dielectric response [37], separated by 180 nm. The fundamental cavity mode is resonant with the first electronic transition of pentacene at 3.975 eV. An ensemble of 101 pentacene molecules is arranged in the $xy$ plane within a 100 nm region at the cavity center, with their molecular long axes aligned along the cavity polarization $x$. Additional details of the simulation setup are provided in the Methods section. Again, the light–matter coupling is controlled by varying the molecular density parameter $N_M$ over the range $3.37 \times 10^{-4}\text{nm}^{-3}$ to $3.37 \times 10^{-2}\text{nm}^{-3}$ (molar concentrations of $5.6 \times 10^{-4}$M and 0.056M).

Figure 3A reveals the simultaneous activation of multiple vibrational modes, with their corresponding VPE exhibiting a clear dependence on the Rabi frequency. We focus on the ten most responsive modes, whose displacement patterns and frequencies are shown in Supplementary Fig. 6. As in the diatomic and benzene cases, these modes are optically inactive and involve in-plane nuclear motion, consistent with the molecular orientation relative to the cavity field. They are characterized by C–C bond distortions, reflecting the weakening of multiple backbone C–C bonds in the electronically excited state that couples to the cavity and agrees with the geometrical selectivity discussed previously. By analyzing the VPE of each vibrational mode as a function of the Rabi frequency, we again observe a Rabi-resonant behavior for different vibrational modes. Each mode's energy is maximized when it is in resonance with the Rab splitting, as shown in Fig. 3B and C, where we present the results for the most sensitive modes in the range of Rabi frequencies considered in this work. Importantly, the orthogonality of the normal modes ensures that each vibrational mode is activated independently, provided the nuclear dynamics in the cavity can be expressed in the normal-mode basis. As discussed above, the peak broadening depends on the cavity quality, while the peak height is determined by the electron–nuclear interaction and the mode's contribution to the cavity nuclear dynamics.

## Discussion

We identify a previously unreported mechanism of cavity-mediated vibrational excitation, which we term Rabi-driven vibrational activation, occurring when molecular ensembles under ESC are driven out of equilibrium. Despite its similarity to previous studies where vibrational activation occurs under VSC conditions [47, 48], we observe that the presented phenomenon exhibits a pronounced maximum when the electronic Rabi frequency resonates with a molecular vibrational frequency, showing a pronounced maximum when the electronic Rabi frequency resonates with a molecular vibrational frequency. Nuclear Ehrenfest dynamics captures this behavior, establishing the Maxwell–DFTB framework as a versatile and predictive approach for studying Rabi-driven vibrational activation in polyatomic molecules. Using benzene and pentacene, we demonstrate that the process is mode-selective and governed by the collective nuclear response to cavity-mediated electronic dynamic. When multiple modes are activated, their orthogonality ensures that each mode is driven independently once its resonance condition is satisfied. Increasing cavity losses broaden lineshapes, so that an exact resonance between the Rabi frequency and the vibrational frequency is somewhat relaxed, enabling the activation of additional vibrational modes. Other sources of broadening, such as disorder, may also affect the process discussed here and deserve to be studied separately.

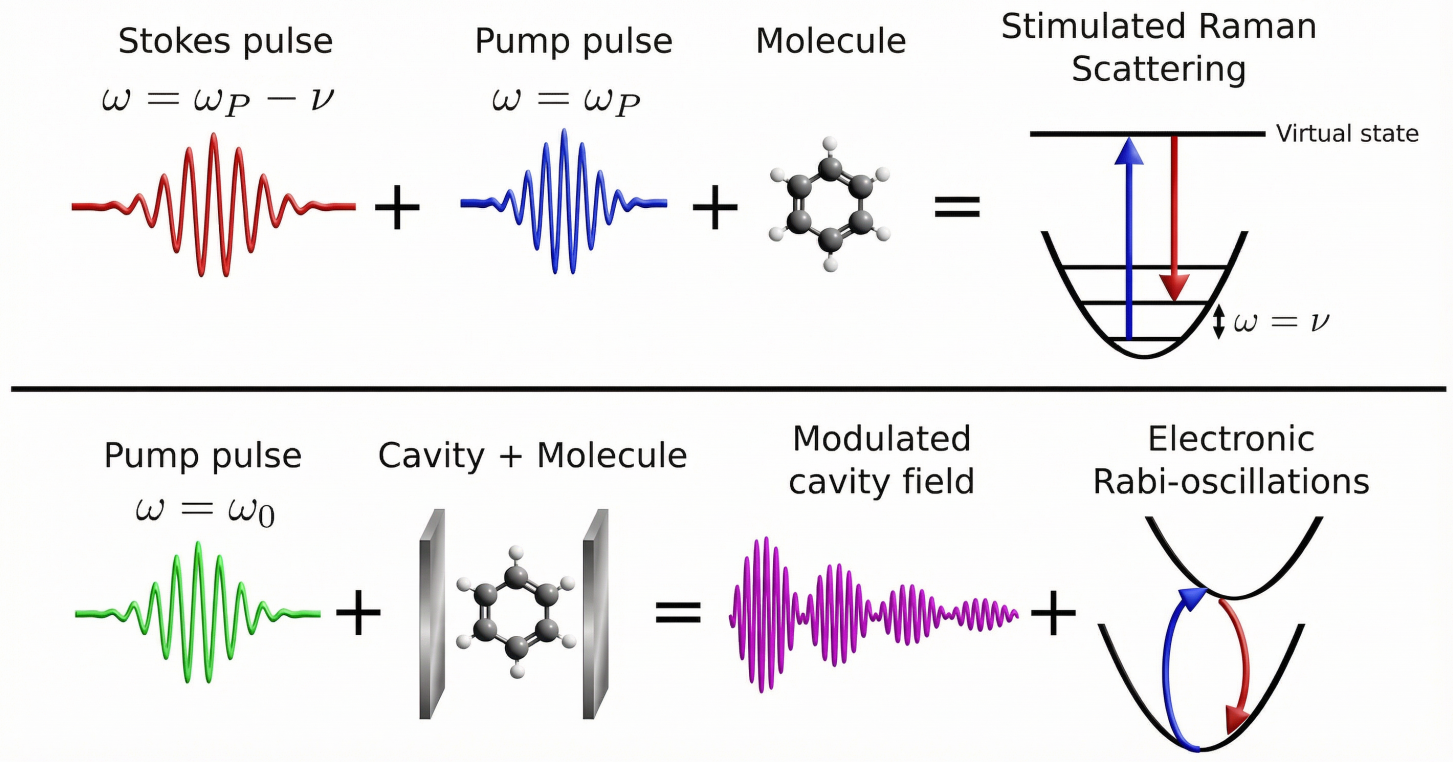

Figure 4: The upper panel illustrates the pump and Stokes pulses, together with their respective frequencies, required to induce stimulated Raman scattering (SRS). In this process, the molecule is promoted to a virtual electronic state, and the subsequent transition to the first vibrational level of the electronic ground state is driven by the Stokes pulse. The lower panel shows the resulting modulated intracavity electromagnetic field and the associated electronic Rabi oscillations that arise following excitation of the coupled cavity–molecule system, under resonant conditions where the cavity mode and the molecular electronic transition satisfy $\omega = \omega_0$.

From a classical perspective, the mechanism can be rationalized in terms of a stimulated Raman-like process. In conventional stimulated Raman scattering (SRS), efficient vibrational excitation occurs when the frequency difference between pump and Stokes pulses matches a vibrational frequency (upper panel of Fig. 4), with the vibrational amplitude scaling with the product of the field amplitudes [49].

In the present simulations, the molecular system is driven by a temporally modulated intracavity field formed by the superposition of the UP and LP fields (lower panel of Fig. 4). A classical analysis (Supplementary Discussion 1) yields an analogous dependence on the product of polaritonic field components, closely mirroring SRS. Unlike conventional SRS, the effective “pump” and “Stokes” fields emerge self-consistently from collective strong light–matter coupling when the system is driven by a short pulse, rather than two independent laser sources. In addition, the intracavity excitation leads to multiple cycles of absorption and emission during the cavity lifetime, in contrast to the single absorption–emission event characteristic of standard SRS (Fig. 4).

An additional SRS-like signature of the Rabi-driven vibrational activation is the quadratic dependence of the population of the $v_1$ vibrational state of the electronic ground state on the electric field intensity (Supplementary Fig. 3), indicative of a second-order nonlinear process (Supplementary Discussion 1). These parallels with Raman scattering provide a rationale for the success of Ehrenfest dynamics in capturing the phenomenon, as this approach has been previously employed to describe Raman spectra and related vibrational processes [36, 50].

From a quantum-electrodynamic perspective, the process corresponds to UP→LP relaxation via phonon emission, thereby inducing vibrational activation. This interpretation is supported by calculations based on the Holstein–Tavis–Cummings (HTC) Hamiltonian (Supplementary Discussion 2), which provide direct access to the UP decay dynamics and the associated phonon generation (Supplementary Fig. 7). In particular, phonon emission is maximized when the Rabi splitting between the UP and LP matches the vibrational frequency.

The semiclassical simulations reproduce this behavior under long-pulse excitation resonant with the UP, although efficient phonon emission in this case still requires the presence of the LP field. As a result, mirror losses and laser linewidth play a critical role in enabling the UP→LP relaxation pathway. Broader cavity modes facilitate LP activation, as does a broader laser spectrum. In contrast, in the limiting case of ideal mirrors or an ultranarrow laser linewidth, phonon emission is fully suppressed. Notably, the HTC model predicts a UP→LP relaxation rate linear in laser intensity, assuming that the initial UP population scales linearly with the laser intensity. This behavior does not reproduce the quadratic dependence of the $v_1$ ground-state population on the electric field intensity observed in the semiclassical simulations.

The HTC model further predicts a pronounced suppression of vibrational energy when the Rabi frequency exceeds the vibrational frequency, in agreement with the behavior observed in our Ehrenfest-based dynamics (Supplementary Fig. 8A). At higher Rabi frequencies, a second maximum appears when the Rabi splitting twice the vibrational frequency (Supplementary Fig. 8B). This feature originates from a two-step relaxation pathway involving an intermediate dark state, followed by relaxation to the LP. Although this second peak resembles the secondary maximum observed in the two-level model results (Fig. 1D), the semiclassical formulation does not explicitly resolve dark states and therefore cannot capture this quantum mechanism at a microscopic level. A detailed analysis of the origin of higher-order peaks in the vibrational populations predicted by the two-level model will be addressed in future work.

Finally, we suggest that this phenomenon could be experimentally explored in gas-phase molecular systems embedded in FP cavities, where the pressure provides a convenient handle to tune the collective Rabi splitting. Alternatively, solid-state platforms based on J-aggregate slabs inside FP cavities offer additional control, as the Rabi splitting can be adjusted by varying the slab thickness or by using angle-resolved excitation. We hope that these considerations will motivate future experimental efforts to investigate the mechanism proposed here.

# Methods

## Maxwell+quantum dynamics

A detailed description of our implementation combining the numerical solution of Maxwell's equations with either nuclear wavepacket dynamics or atomistic electronic dynamics based on density functional tight binding (DFTB) can be found in Refs. [42, 51] and Ref. [33], respectively. Here, we summarize the key elements relevant to the present work.

We solve Maxwell's equations for the electric field $\mathbf{E}$ and magnetic field $\mathbf{B}$ on a spatial grid using the finite-difference time-domain (FDTD) method [52],

$$\begin{aligned}\frac{\partial \mathbf{B}(\mathbf{r},t)}{\partial t} &= -\boldsymbol{\nabla} \times \mathbf{E}(\mathbf{r},t),\\ \frac{\partial \mathbf{E}(\mathbf{r},t)}{\partial t} &= c_0^2 \boldsymbol{\nabla} \times \mathbf{B}(\mathbf{r},t) - \frac{1}{\epsilon_0}\mathbf{J}(\mathbf{r},t),\end{aligned} \tag{1}$$

where $\mathbf{J}$ is the total current density. Simulations are performed in one and two spatial dimensions. Open boundary conditions are implemented using the convolutional perfectly matched layer (CPML) method [52].

The current density $\mathbf{J}$ includes contributions from both the cavity mirrors and the molecular polarization. The optical response of the metallic mirrors is described using a multi-pole Drude–Lorentz dielectric function,

$$\varepsilon(\omega) = 1 - \frac{\Omega_D^2}{\omega^2 - \mathrm{i}\Gamma_D\omega} - \sum_n \frac{\Delta\varepsilon_n \omega_p^2}{\omega^2 - \omega_n^2 - \mathrm{i}\Gamma_n\omega}, \tag{2}$$

with material parameters taken from Ref. [37]. Aluminium parameters are used over the energy range 0.01–10 eV and gold parameters over 0.2–5.0 eV, ensuring an accurate description of realistic cavity dispersion.

The dielectric response in Eq. 2 is implemented through auxiliary differential equations for the metal current densities,

$$\frac{\partial \mathbf{J}_D(\mathbf{r},t)}{\partial t} + \Gamma_D \mathbf{J}_D(\mathbf{r},t) = \varepsilon_0 \Omega_D^2 \mathbf{E}(\mathbf{r},t), \tag{3}$$

$$\frac{\partial^2 \mathbf{J}_n(\mathbf{r},t)}{\partial t^2} + \Gamma_n \frac{\partial \mathbf{J}_n(\mathbf{r},t)}{\partial t} + \omega_n^2 \mathbf{J}_n(\mathbf{r},t) = \varepsilon_0 \Delta\varepsilon_n \omega_p^2 \frac{\partial \mathbf{E}(\mathbf{r},t)}{\partial t}. \tag{4}$$

The molecular contribution to the current density is given by $\mathbf{J}_{\mathrm{mol}} = \partial \mathbf{P}_{\mathrm{mol}}/\partial t$, where $\mathbf{P}_{\mathrm{mol}}$ is the macroscopic molecular polarization. The polarization at the position $\mathbf{r}_A$ of molecule $A$ is computed as

$$\mathbf{P}(\mathbf{r}_A, t) = N_M \langle \hat{\boldsymbol{\mu}}_A \rangle, \tag{5}$$

where $N_M$ is the molecular number density and $\langle \hat{\boldsymbol{\mu}}_A \rangle$ is the expectation value of the molecular dipole moment [42, 51, 53]. Each molecule is assigned to an individual grid point, forming an effective continuous macroscopic medium. The molecular dipole moments are obtained from the time propagation of the electronic density matrix [33, 36].

## Two-level model with vibrations

As a minimal and exacty solvable molecular model, we propagate the time-dependent Schrödinger equation using a two-level Hamiltonian coupled to nuclear motion,

$$\hat{H} = \begin{bmatrix} \hat{T} & 0 \\ 0 & \hat{T} \end{bmatrix} + \begin{bmatrix} V_g & -\mu_{eg} E_x \\ -\mu_{eg} E_x & V_e \end{bmatrix}, \tag{6}$$

where $\hat{T}$ is the nuclear kinetic energy operator, $\mu_{eg}$ is the electronic transition dipole moment, and $E_x$ is the $x$-component of the electric field, corresponding to the polarization used in the one-dimensional simulations.

The ground- and excited-state potential energy surfaces are modeled as harmonic oscillators,

$$\begin{aligned} V_g &= \frac{1}{2} M\nu (R - R_0)^2, \\ V_e &= \frac{1}{2} M\nu (R - R_0 - \Delta R)^2 + \Delta V, \end{aligned} \tag{7}$$

where $M$ is the reduced mass, $R_0$ is the ground-state equilibrium coordinate, $\Delta R$ is the shift between ground- and excited-state equilibria, and $\Delta V$ is the electronic excitation energy. We use parameters corresponding to an $N_2$-like reduced mass, with $\Delta V = 1.6985$ eV, $\Delta R = 0.023$ a.u., and $R_0 = 0$. Unless stated otherwise, the vibrational frequency is $\nu = 0.10$ eV. Additional values $\nu = 0.08$ eV and 0.11 eV are used for Fig. 1.

The nuclear wavepacket dynamics is propagated using the split-operator method with a time step of 0.05 fs. Each molecule acts as an induced dipole under the local electric field, and its time derivative contributes to the Maxwell equations via Eq. 5. The molecular density is fixed at $N_M = 7.0 \times 10^{-2}$ nm$^{-3}$. The system is excited by a short pulse with peak amplitude 0.5 V/nm and duration 5 fs. FDTD grid spacing and time steps match those used in the DFTB simulations.

## DFTB systems

Benzene and pentacene molecules are simulated using time-dependent density functional tight binding (TDDFTB) theory [54, 55], as implemented in the DFTB+ package [35, 36]. Nuclear motion is treated at the Ehrenfest level [36]. The mio-1-1 Slater–Koster parameter set is employed [54], and molecular geometries are optimized prior to the coupled Maxwell–TDDFTB simulations.

For benzene in one-dimensional cavities, we use a grid spacing $\Delta z = 1.0$ nm, a Maxwell time step $\Delta t_{\mathrm{Mxll}} = 2.419 \times 10^{-4}$ fs, a molecular time step $\Delta t_{\mathrm{mol}} = 5\Delta t_{\mathrm{Mxll}}$, and a total simulation time of 400 fs. The excitation pulse has peak amplitude 0.05 V/nm, frequency 7.0 eV, and duration 4.0 fs.

Pentacene simulations use a one-dimensional cavity with $\Delta z = 1.0$ nm, identical time steps, and a total simulation time of 400 fs. The excitation pulse has peak amplitude 0.7 V/nm, frequency 4.0 eV, and duration 6.0 fs.

## Average vibrational potential energy

To quantify vibrational activation, nuclear displacements are projected onto the molecular normal modes [56, 57],

$$Q_i(t) = \sum_A m_A \Delta \mathbf{r}_A(t) \cdot \mathbf{v}_{A,i}, \tag{8}$$

where the sum runs over nuclei $A$, $\mathbf{v}_{A,i}$ are the normal mode eigenvectors, $\Delta\mathbf{r}_A$ are displacements relative to equilibrium, and $m_A$ are atomic masses.

The potential energy associated with mode $i$ is then defined as

$$V_i(t) = \frac{1}{2}\left(\nu_i Q_i(t)\right)^2 , \tag{9}$$

where $\nu_i$ is the vibrational frequency of the mode. The average vibrational potential energy (VPE) is obtained by time averaging $V_i(t)$ over the final portion of the simulation, after most of the optical energy has dissipated. For benzene and pentacene, averages are taken over the final 100 fs.

## Acknowledgements

Funded by the European Union under the ERC Synergy Grant UnMySt (HEU GA No. 101167294). Views and opinions expressed are however those of the author(s) only and do not necessarily reflect those of the European Union or the European Research Council. Neither the European Union nor the European Research Council can be held responsible for them. This work was also supported by the Cluster of Excellence Advanced Imaging of Matter (AIM), Grupos Consolidados (IT1249-19) and SFB925. We acknowledge support from the Max Planck-New York City Center for Non-Equilibrium Quantum Phenomena. The Flatiron Institute is a division of the Simons Foundation. C. M. Bustamante thanks the Alexander von Humboldt-Stiftung for the financial support from the Humboldt Research Fellowship. M.S. acknowledges support by the Office of Naval Research, Grant No. N000142512090 and the Air Force Office of Scientific Research under Grant No. FA9550-25-1-0096. F.P.B. acknowledges financial support from the European Union's Horizon 2020 research and innovation program under the Marie Sklodowska-Curie Grant Agreement no. 895747 (NanoLightQD).

## Supporting information

The following files are available free of charge.

- Supplementary movie: shows selected observables from Maxwell–DFTB dynamics of benzene molecules in a Fabry–Pérot cavity under electronic strong coupling.
- Supplementary information: additional results and discussions supporting the main text, as well as a more detailed caption for the Supplementary Movie.

# Supplementary Information for Collective Rabi-driven vibrational activation in molecular polaritons

Carlos M. Bustamante [*,1], Franco P. Bonafé [†,1], Richard Richardson[2], Michael Ruggenthaler[1], Wenxiang Ying[3], Abraham Nitzan [‡,3], Maxim Sukharev [§,2,4], and Angel Rubio [¶,1,5]

[1]Max Planck Institute for the Structure and Dynamics of Matter and Center for Free-Electron Laser Science, Luruper Chaussee 149, Hamburg 22761, Germany
[2]Department of Physics, Arizona State University, Tempe, Arizona 85287, United States
[3]Department of Chemistry, University of Pennsylvania, Philadelphia, Pennsylvania 19104, United States
[4]College of Integrative Sciences and Arts, Arizona State University, Mesa, Arizona 85212, United States
[5]Initiative for Computational Catalysis (ICC), Flatiron Institute, Simons Foundation, 162 5th Avenue, New York, NY 10010 USA

*Throughout this Supplementary Information, the symbol $\Omega$ denotes the Rabi splitting between the upper and lower polariton modes expressed in energy units, $\nu$ denotes molecular vibrational mode energies, and $N_M$ denotes the effective molecular number density used to control the collective light–matter coupling strength.*

## Supplementary Results

*Supplementary Figure 1* presents results obtained with the two-level model when different mirror thicknesses are considered. Decreasing the mirror thickness increases cavity losses. The final $v_1$ population, plotted versus $\mu_{eg}$, still exhibits a pronounced maximum, but the resonance peak broadens as losses increase. Changes in mirror thickness also shift the cavity mode structure; to maintain resonance with the molecule we adjusted molecular parameters accordingly, which in turn required different values of $\mu_{eg}$ to reach resonance.

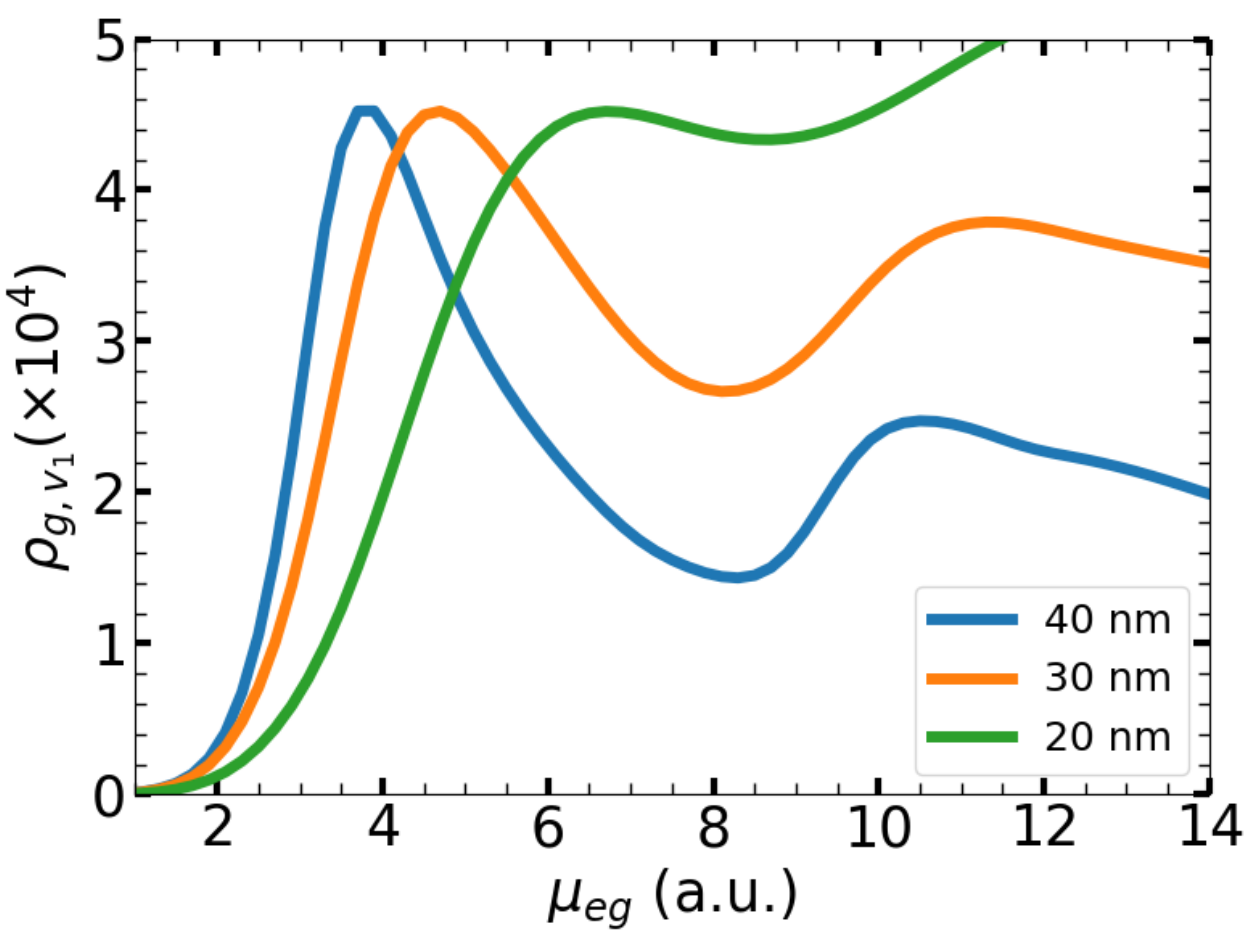

Supplementary Figure 1: Occupation of the first vibrational state $v_1$ of the electronic ground state at long times, plotted as a function of $\mu_{eg}$ for three mirror thicknesses.

*carlos.bustamante@mpsd.mpg.de
†franco.bonafe@mpsd.mpg.de
‡anitzan@sas.upenn.edu
§Maxim.Sukharev@asu.edu
¶angel.rubio@mpsd.mpg.de

*Supplementary Figure 2* shows that, for larger molecular slab sizes, both the value of $\mu_{eg}$ required to reach the resonant condition and the maximum vibrational population are reduced.

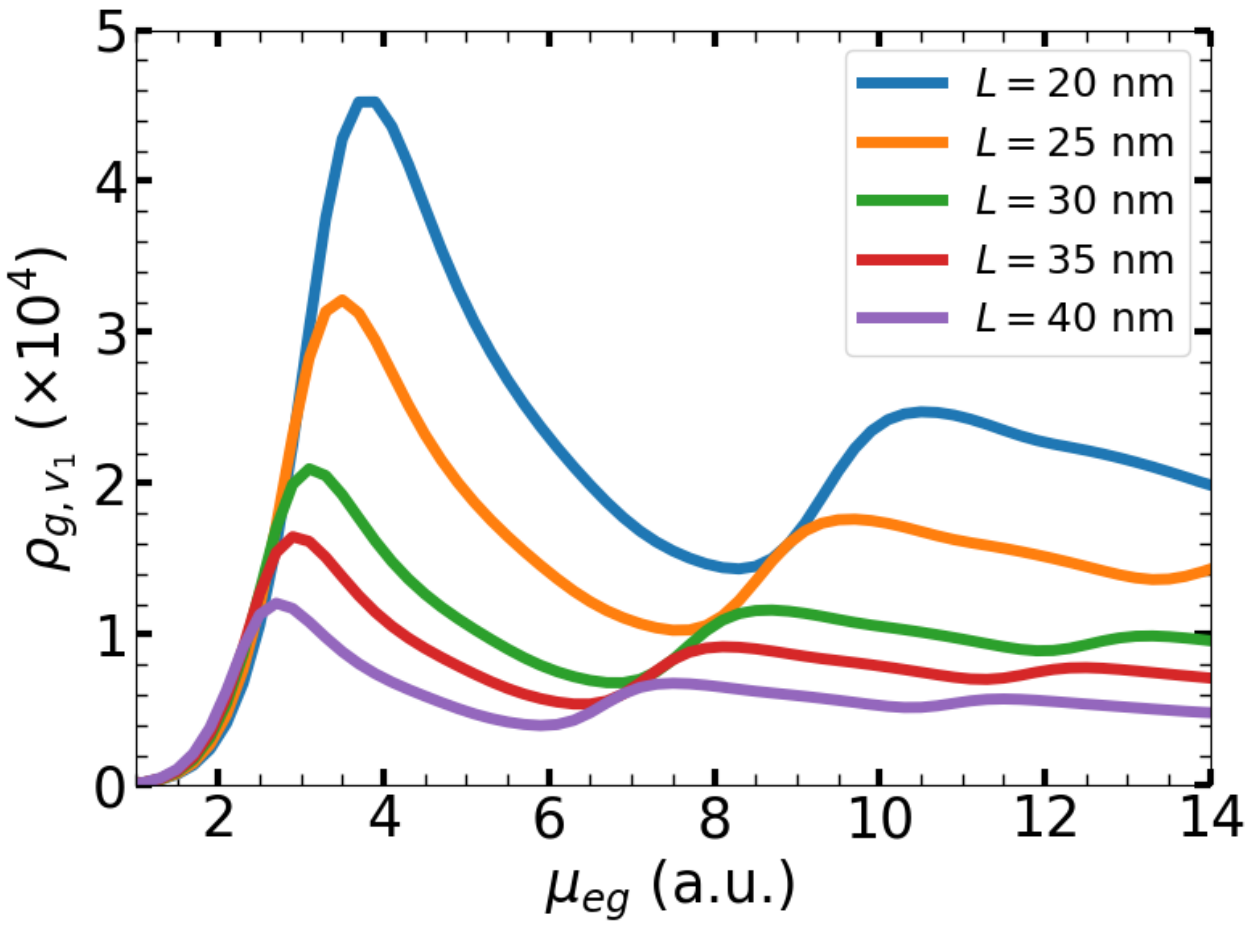

Supplementary Figure 2: Occupation of the first vibrational state $v_1$ of the electronic ground state at long times, plotted as a function of $\mu_{eg}$ for different molecular slab sizes ($L$).

*Supplementary Figure 3* shows the vibrational population at resonance as a function of the applied electric field amplitude. The fit indicates $\rho_{g,v_1} \propto |E_x|^4$, equivalent to $\rho_{g,v_1}$ scaling with the square of the laser intensity, consistent with a second-order process.

*Supplementary Figure 4* shows transmission spectra from Ehrenfest simulations of the benzene ensemble described in the main text. The two polaritonic peaks are readily identified, allowing extraction of the Rabi splitting $\Omega$ and the corresponding spectral Rabi frequency.

*Supplementary Figure 5* shows the results of the extended analysis of Rabi-driven vibrational activation to two-dimensional. The simulation box spans $340 \times 340$ nm$^2$ in the $xy$ plane with grid spacing $\Delta x = \Delta y = 1$ nm. A 20 nm boundary region is assigned to the CPML. Time steps and total simulation time are identical to the one-dimensional case. The system is excited by a point-like pulse applied at the cavity center with frequency 7.0 eV and a Gaussian envelope of 1.5 fs full width at half maximum. In this case, we consider a ring cavity geometry, as illustrated in Supplementary Fig. 5A, with an inner radius of 56 nm and an outer radius of 106 nm. The fundamental cavity mode is tuned to resonance with the first electronic transition of the benzene molecule. As in the one-dimensional case, the optical response of the cavity mirrors is described using Drude–Lorentz parameters for aluminium. A total of 221 benzene molecules are placed at the center of the cavity on the circumference of a circle with a radius of 8 nm, with each molecule occupying an individual grid point, as shown in the inset of Supplementary Fig. 5A. The strength of the light–matter coupling is controlled by varying the molecular density parameter $N_M$ between 0.067 nm$^{-3}$ and 0.337 nm$^{-3}$, corresponding to molar concentrations of 0.11 M and 0.56 M, respectively.

Supplementary Fig. 5B shows how the higher losses introduced by the additional spatial dimension decrease the specificity of Rabi-driven vibrational activation, leading to the concurrent activation of multiple vibrational modes for a given Rabi splitting.

Despite the increased broadening of the resonant response due to these losses, the hallmark of Rabi-driven vibrational activation remains clearly observable in the two-dimensional cavity, as shown in Supplementary Fig. 5C. In particular, the VPE of the benzene breathing mode retains a pronounced dependence on the Rabi frequency and exhibits a maximum under resonant conditions.

*Supplementary Figure 6* displays the atomic-displacement patterns of the normal modes in pentacene that are most sensitive to the Rabi-driven effect discussed in Figure 4 of the main text. These modes are optically inactive by symmetry and are characterized by significant C–C displacements.

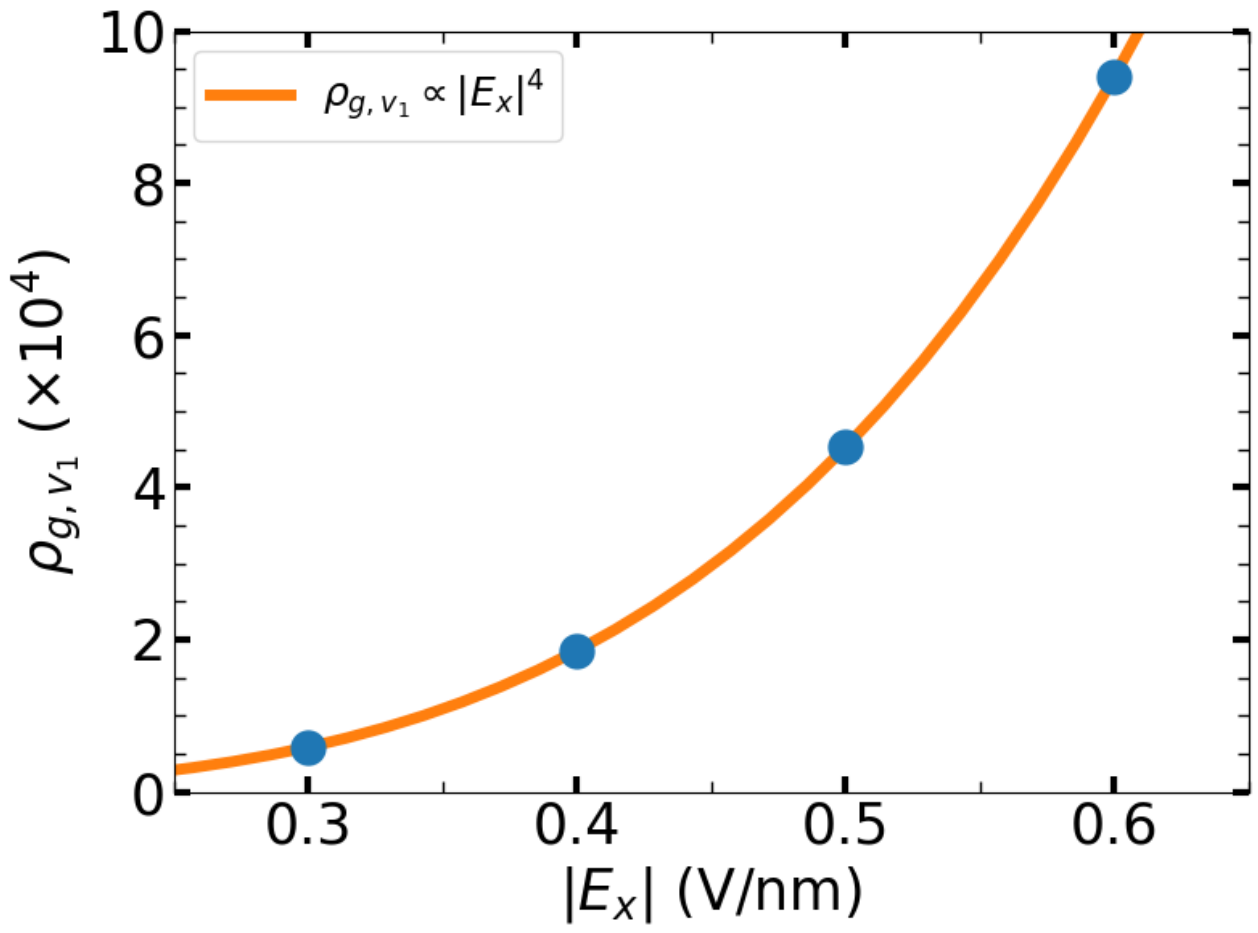

Supplementary Figure 3: Occupation of the first vibrational state $v_1$ of the ground state at resonance ($\nu = 0.1$ eV) as a function of the peak amplitude of the source electric field (blue dots). The orange curve is a fit showing a dependence $\rho_{g,v_1} \propto |E_x|^4$.

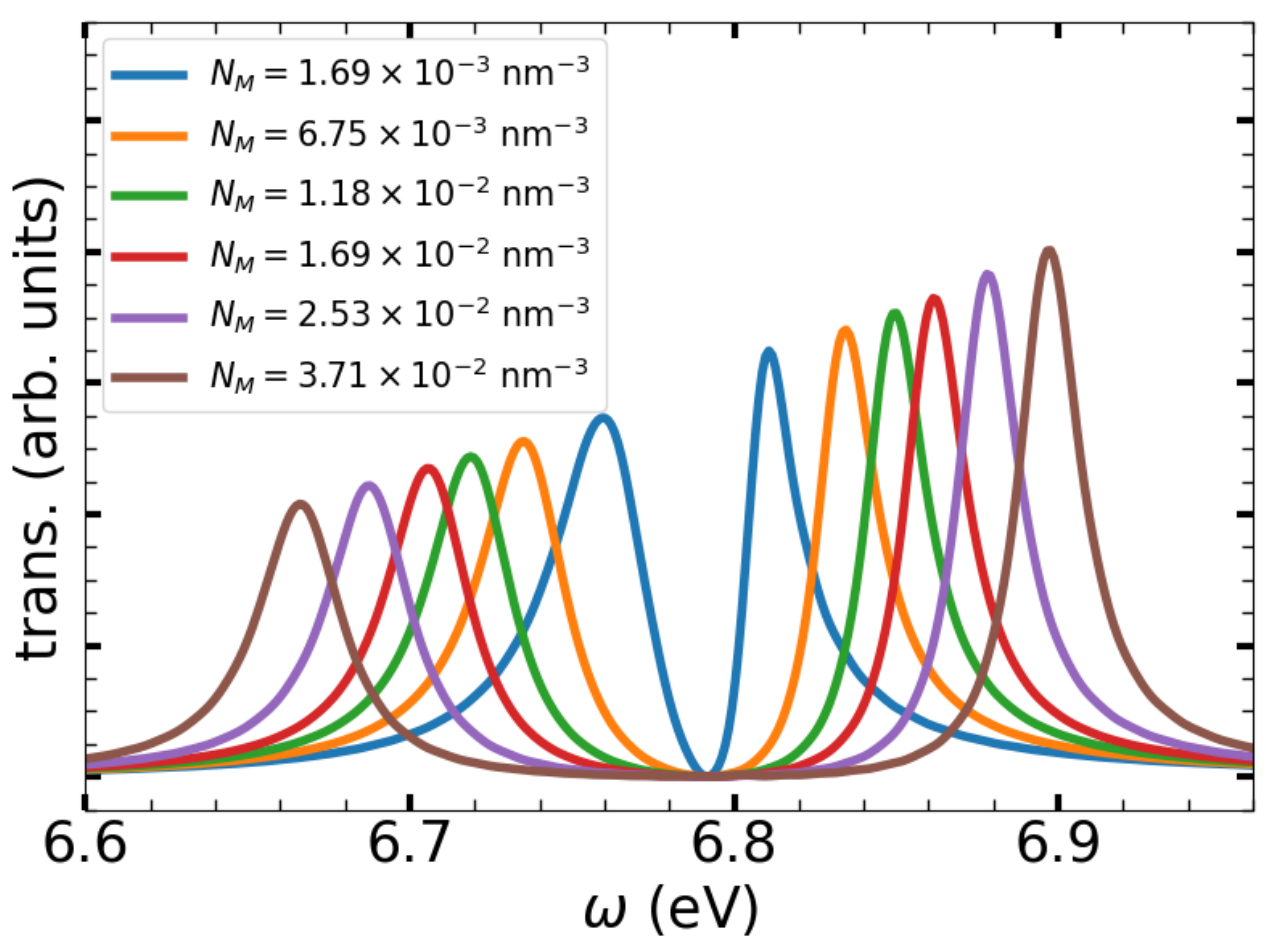

Supplementary Figure 4: Transmission spectra for 201 benzene molecules inside an optical cavity resonant with the first electronic transition, shown for different values of the density parameter $N_M$.

# Supplementary Discussion 1: Classical forced oscillator / SRS analogy

A classical picture provides a transparent connection between the Rabi-driven activation described in the main text and a stimulated Raman-like mechanism. Consider a forced, damped harmonic oscillator describing a normal coordinate $Q$,

$$\ddot{Q} + \gamma \dot{Q} + \nu^2 Q = F_Q(t), \tag{S1}$$

where $\nu$ is the vibrational frequency, $\gamma$ is a damping rate, and $F_Q(t)$ is the driving force. Assuming a Raman-like origin for the driving force,

$$F_Q(t) \approx \tfrac{1}{2} \frac{\partial \alpha}{\partial Q} |E(t)|^2, \tag{S2}$$

with $\alpha$ the dynamic polarizability and $E(t)$ the electric field. Writing the intracavity field as a superposition of lower- and upper-polariton components,

$$E(t) = E_{LP} e^{-\mathrm{i}\omega_{LP} t} + E_{UP} e^{-\mathrm{i}\omega_{UP} t} + \text{c.c.}, \tag{S3}$$

the intensity contains a beat term,

$$|E(t)|^2 \supset 2|E_{LP}||E_{UP}| \cos(\Omega t), \tag{S4}$$

where $\Omega = \omega_{UP} - \omega_{LP}$ is the Rabi splitting. (Other terms oscillate at optical frequencies and average out for $\nu \ll \omega_{LP}, \omega_{UP}$.)

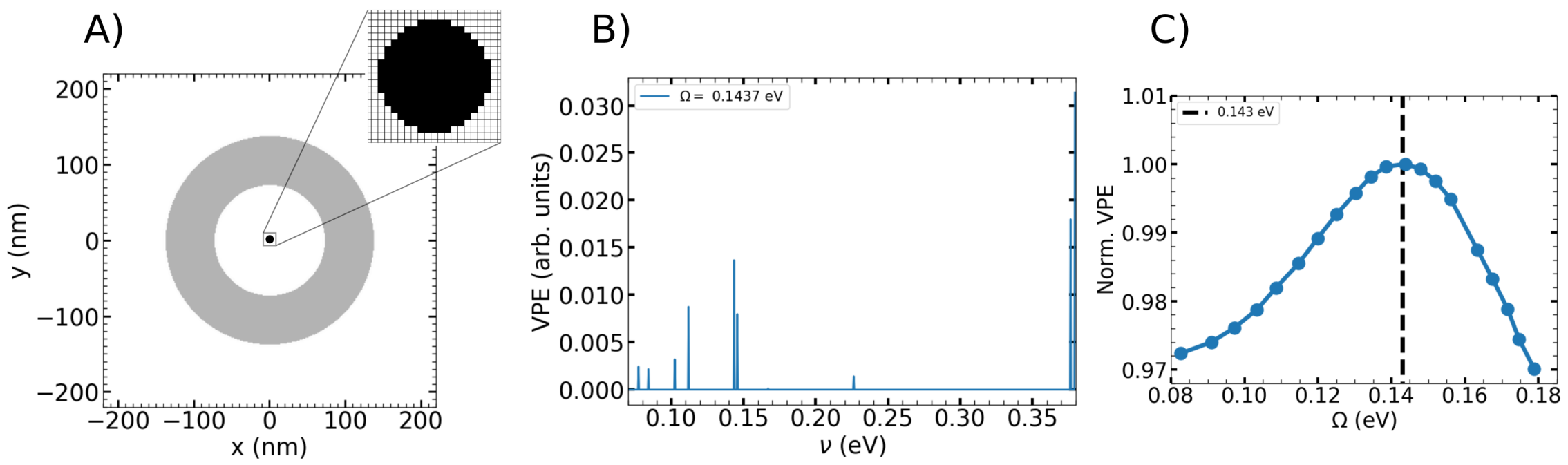

Supplementary Figure 5: A) Schematic representation of the ring cavity used in the two-dimensional simulations. The cavity has an inner radius of 56 nm and an outer radius of 106 nm. A total of 221 benzene molecules are placed at the center of the cavity, occupying a circular region with a radius of 8 nm. The inset highlights the grid points corresponding to the positions of individual molecules. B) VPE of the central benzene molecule, located at $\mathbf{r} = (0.0, \text{nm}, 0.0, \text{nm})$, plotted as a function of vibrational frequency and averaged over the final 100 fs of the simulation. C) Normalized VPE of the benzene breathing mode of the central molecule as a function of the Rabi splitting extracted from the transmission spectra, obtained by varying the molecular density parameter $N_M$.

Inserting the beat term into Eq. (S1) and solving for the long-time driven amplitude yields

$$Q_0 \propto |\chi_\nu(\Omega)|\,|E_{LP}|\,|E_{UP}|, \tag{S5}$$

with the susceptibility

$$\chi_\nu(\Omega) \propto \frac{1}{\nu^2 - \Omega^2 - \mathrm{i}\gamma\Omega}. \tag{S6}$$

Thus the driven amplitude is maximal when $\Omega \approx \nu$, reproducing the resonance condition observed in the simulations. The combination of Eqs. (S5) and (S6) parallels the dependence found in stimulated Raman scattering, with $E_{LP}$ and $E_{UP}$ playing roles analogous to pump and Stokes fields [1].

Assuming that only the vibrational states $v_0$ and $v_1$ are significantly populated and that $\rho_{v_1} \ll \rho_{v_0}$, one obtains $\langle Q \rangle \propto \sqrt{\rho_{v_1}}$. Combining this relation with Eq. (S5) explains the observed scaling $\rho_{g,v_1} \propto |E_x|^4$ reported in Supplementary Fig. 3.

# Supplementary Discussion 2: Holstein–Tavis–Cummings analysis

We present here a Holstein–Tavis–Cummings (HTC) model [2, 3] to analyze Rabi-driven vibrational excitation from a quantum-electrodynamical perspective. The HTC model describes $N$ electronic two-level systems, each coupled to a local harmonic vibration, and collectively coupled to a single cavity mode. The HTC analysis complements the mean-field Maxwell+Ehrenfest simulations: the latter capture spatially resolved macroscopic polarization and nonlinear response but neglect light–matter entanglement, while the HTC model resolves bright and dark manifolds and phonon-mediated relaxations in the single-excitation regime.

The HTC Hamiltonian reads

$$\hat{H}_{\mathrm{HTC}} = \hat{H}_{\mathrm{M}} + \hat{H}_{\mathrm{cav}} + \hat{H}_{\mathrm{LM}}, \tag{S7}$$

with

$$\hat{H}_{\mathrm{M}} = \sum_{n=1}^{N} \hbar\omega_0 \hat{\sigma}_n^+ \hat{\sigma}_n^- + \hbar\nu \sum_n \hat{b}_n^\dagger \hat{b}_n + \sum_n \hat{\sigma}_n^+ \hat{\sigma}_n^- \, c_\nu (\hat{b}_n + \hat{b}_n^\dagger), \tag{S8}$$

$\hat{H}_{\mathrm{cav}} = \hbar\omega_{\mathrm{c}}(\hat{a}^\dagger \hat{a} + \frac{1}{2})$, and $\hat{H}_{\mathrm{LM}} = \hbar g_{\mathrm{c}} \sum_{n=1}^{N} (\hat{a}^\dagger \hat{\sigma}_n^- + \hat{a}\hat{\sigma}_n^+)$. Here $\hat{\sigma}_n^\pm$ are excitonic raising/lowering operators, $\hat{b}_n$ are local phonon annihilation operators of frequency $\nu$, and $c_\nu$ is the exciton–phonon coupling. The shift between ground- and excited-state minima is $\Delta R = \sqrt{2c_\nu^2/\nu^3}$ and the Huang–Rhys factor is $S = (c_\nu/\nu)^2$.

The purely electronic-photonic part forms the Tavis–Cummings (TC) Hamiltonian,

$$\hat{H}_{\mathrm{TC}} = \sum_{n=1}^{N} \hbar\omega_0 \hat{\sigma}_n^+ \hat{\sigma}_n^- + \hbar\omega_{\mathrm{c}} \left(\hat{a}^\dagger \hat{a} + \tfrac{1}{2}\right) + \hbar g_{\mathrm{c}} \sum_{n=1}^{N} (\hat{a}^\dagger \hat{\sigma}_n^- + \hat{a}\hat{\sigma}_n^+). \tag{S9}$$

In the single-excitation manifold one can define the collective bright state $|\mathrm{B}\rangle = \frac{1}{\sqrt{N}} \sum_n |\mathrm{E}_n, 0\rangle$ which couples to the one-photon state $|\mathrm{G}, 1\rangle$. Diagonalization yields the upper and lower polariton eigenstates $|\pm\rangle$

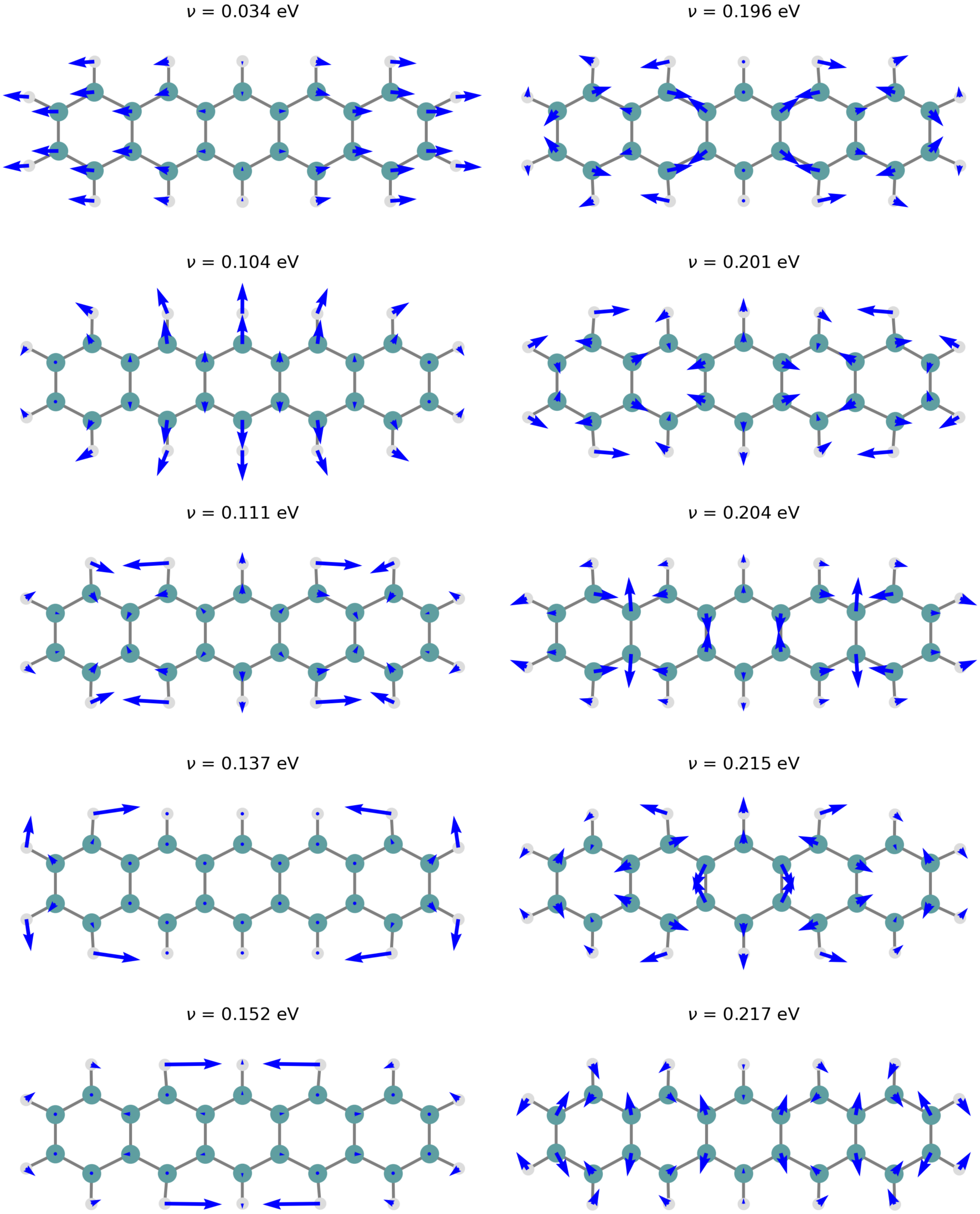

Supplementary Figure 6: Atomic displacement patterns for the ten normal modes most strongly affected by the Rabi oscillations, together with their vibrational frequencies.

with energies $\omega_{\pm}$. We define the collective Rabi splitting

$$\Omega \equiv \omega_+ - \omega_- = \sqrt{(\omega_c - \omega_0)^2 + 4Ng_c^2}, \tag{S10}$$

which reduces to $\Omega = 2\sqrt{N}g_c$ at resonance $\omega_c = \omega_0$.

Dark states $|D_k\rangle$ (with $k = 1, \ldots, N-1$) complete the single-excitation basis; their energies remain at the exciton site energy in the homogeneous model. Fourier transforming the phonon operators, $\hat{b}_k = \frac{1}{\sqrt{N}} \sum_n e^{2\pi i n k/N}\, \hat{b}_n$, and expressing the HTC Hamiltonian in the polariton basis yields exciton–phonon coupling terms that mediate transitions among $|+\rangle$, $|-\rangle$, and $|D_k\rangle$. In particular, transitions become resonant when energy gaps match phonon energies.

We solve the time-dependent Schrödinger equation,

$$\hat{H}_{\mathrm{HTC}} |\Psi(t)\rangle = i\hbar \frac{\partial}{\partial t} |\Psi(t)\rangle, \tag{S11}$$

within the single-excitation manifold using a finite vibrational Fock basis for each mode. Simulations are initialized with the system in the upper polariton and all vibrations in their ground state, $|\Psi(0)\rangle = |+\rangle \bigotimes_{n=1}^{N} |0_n\rangle$.

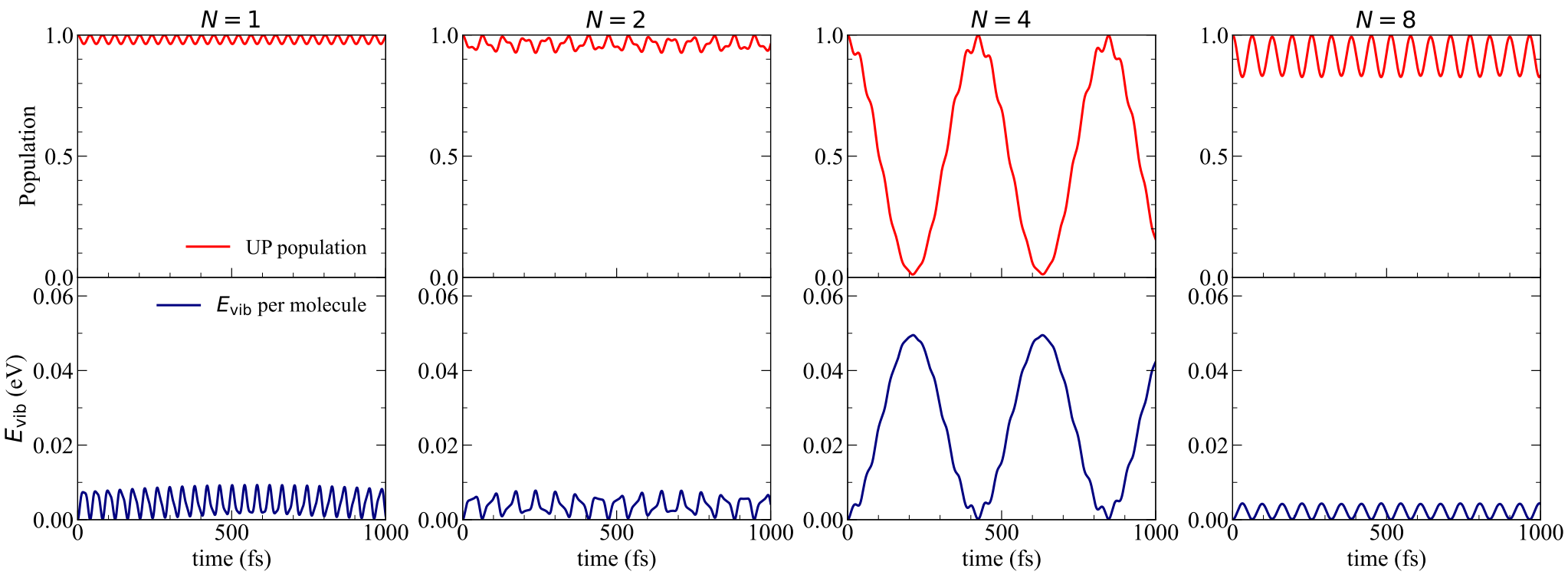

Supplementary Figure 7: Quantum dynamics of the UP population and vibrational energy per molecule for different numbers of molecules $N$.

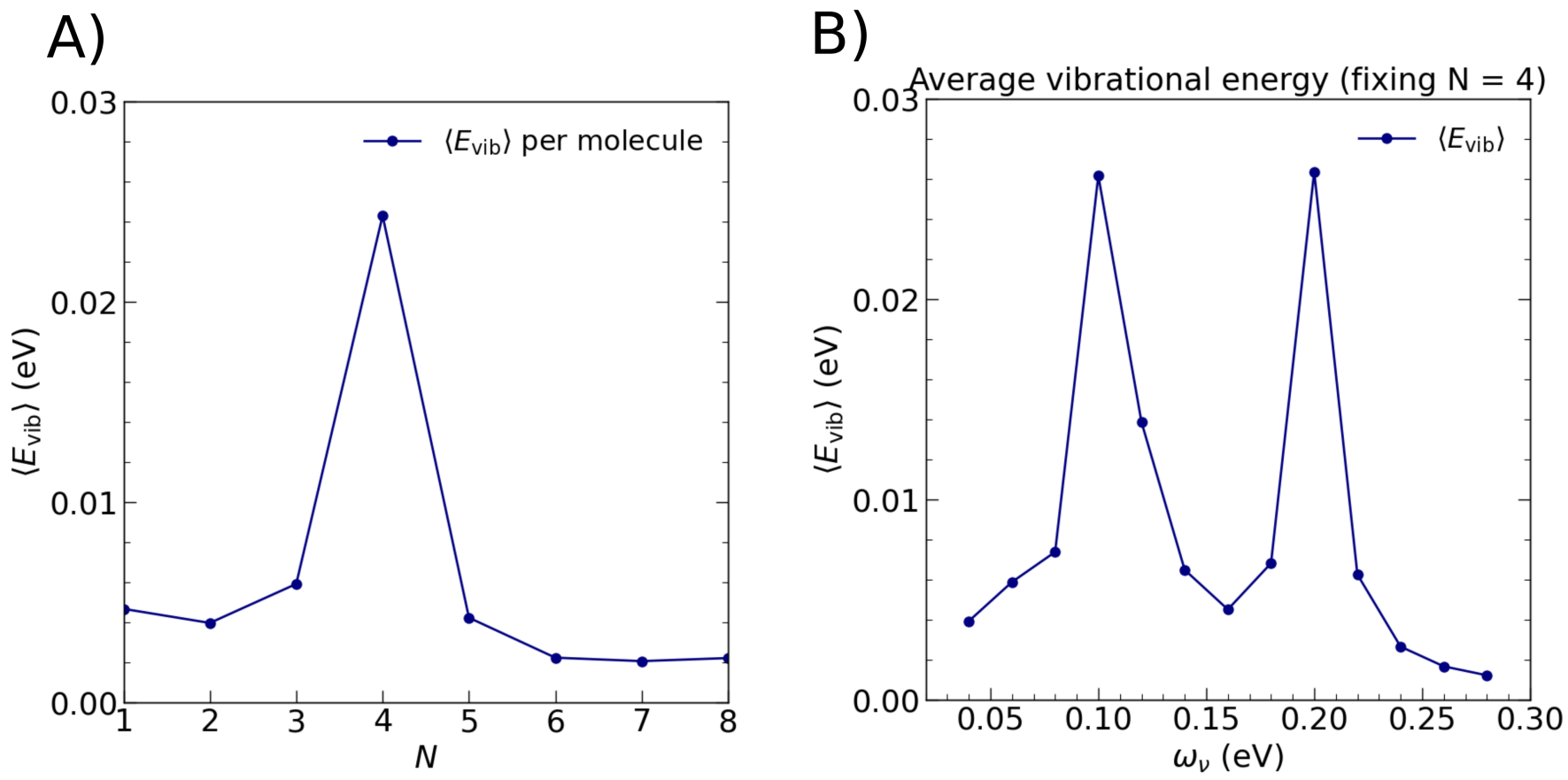

Supplementary Figure 8: Time-averaged vibrational energy per molecule. (A) $\langle E_{\mathrm{vib}} \rangle$ as a function of $N$ for fixed $g_{\mathrm{c}} = 0.05$ eV and $\nu = 0.2$ eV, showing a maximum at resonance $N = 4$. (B) $\langle E_{\mathrm{vib}} \rangle$ as a function of $\nu$ for fixed $N = 4$ and $\Omega = 0.2$ eV, showing resonances at $\nu = \Omega/2$ and $\nu = \Omega$.

The dynamics are propagated with a Trotter integrator with time step $\Delta t = 0.025$ fs and total propagation $T = 1$ ps.

We monitor the UP population

$$P_{\mathrm{UP}}(t) = \langle \Psi(t) | + \rangle \langle + | \Psi(t) \rangle, \tag{S12}$$

and the vibrational energy per molecule

$$E_{\mathrm{vib}}(t) = \frac{\hbar \nu}{N} \sum_{n=1}^{N} \langle \Psi(t) | \hat{b}_n^\dagger \hat{b}_n | \Psi(t) \rangle. \tag{S13}$$

Supplementary Fig. 7 shows representative quantum dynamics of the UP population and vibrational energy per molecule for different values of $N$, with $g_{\mathrm{c}} = 0.05\,\mathrm{eV}$ and $\nu = 0.20\,\mathrm{eV}$. The resonance condition $\nu = \Omega$ is reached for $N = 4$ (since $\Omega = 2\sqrt{N} g_{\mathrm{c}}$ at resonance). Off resonance ($N = 1, 2, 8$) the UP population remains close to unity and the vibrational energy per molecule stays small, while at resonance ($N = 4$) UP population dynamics show pronounced oscillations accompanied by large-amplitude vibrational energy oscillations, indicating efficient polariton-to-phonon energy transfer.

To quantify vibrational excitation efficiency we compute the time-averaged vibrational energy per molecule,

$$\langle E_{\mathrm{vib}} \rangle = \frac{1}{T} \int_0^T dt\, E_{\mathrm{vib}}(t), \tag{S14}$$

with total propagation $T = 1$ ps.

Supplementary Fig. 8 summarizes $\langle E_{\mathrm{vib}} \rangle$ as a function of system parameters. Panel A fixes $g_{\mathrm{c}} = 0.05$ eV and $\nu = 0.2$ eV and scans $N$, revealing a clear maximum at $N = 4$ where $\nu = \Omega$. Panel B fixes $N = 4$ and $\Omega = 0.2$ eV and scans $\nu$, revealing two pronounced resonances at $\nu = \Omega/2$ and $\nu = \Omega$, corresponding to UP $\rightarrow$ DS $\rightarrow$ LP (two-phonon) and UP $\rightarrow$ LP (one-phonon) relaxation channels, respectively.

Taken together, the Maxwell+Ehrenfest and the HTC descriptions identify the same polaritonic normal modes and vibronically induced resonances. In the QED picture, UP $\rightarrow$ LP + vibron(s) transitions describe polariton relaxation into phonons, whereas in the Maxwell+Ehrenfest picture the same physics appears as Raman-like frequency mixing between coupled field and polarization amplitudes. The essential formal distinction is that mean-field factorization in Maxwell+Ehrenfest eliminates light–matter entanglement and represents dark-state physics implicitly via spatially structured polarization modes, allowing for nonlinear molecular responses; the QED HTC model, by contrast, resolves bright and dark manifolds explicitly but is restricted to the single-excitation (linear-response) regime.

## Caption for Supplementary Movie

The movie shows selected observables from Maxwell–DFTB dynamics of benzene molecules in a FP cavity under ESC, when the corresponding Rabi-frequency resonates with the benzene breathing mode. The left panel shows the time evolution of the electric field in the simulation box. The gray rectangles indicate the positions of the aluminum mirrors, and the cyan rectangle marks the region containing 201 benzene molecules. The dynamics is initiated by pumping the system with an ultrashort pulse incident on the cavity through the right mirror. The middle panel shows the nuclear displacement of the benzene molecule located at the center of the cavity. The displacement is amplified by a factor of 5000 to visually highlight the breathing-mode–like nuclear dynamics. The right panel shows the time evolution of the electronic energy of the benzene molecule at the center of the FP cavity.